\documentclass[11pt]{article} 
\usepackage[latin1]{inputenc}
\usepackage[T1]{fontenc} 
\usepackage{fullpage}
\usepackage{latexsym}
\usepackage{epsfig}
\usepackage{amssymb}
\usepackage{array}
\usepackage{float}
\usepackage{algorithm}
\usepackage{algorithmic}
\graphicspath{{figures/}{./}}

\newenvironment{proof}{{\bf Proof. } }{{\hfill $\Box$}\vspace{.5pc}}
\newtheorem{theorem}{Theorem}

\newtheorem{definition}[theorem]{Definition}
\newtheorem{lemma}[theorem]{Lemma}

\floatstyle{ruled}

\floatstyle{ruled}
\newfloat{algo}{thp}{lop}
\floatname{algo}{Algorithm}

\newcommand{\OR}{\mathtt{O}}
\newcommand{\IF}[1]{\textbf{if} {#1}}
\newcommand{\THEN}{\textbf{then} }
\newcommand{\ELSE}{\textbf{else} }

\newcommand{\ENDIF}{\textbf{endif} }
\newcommand{\FOR}[3]{\textbf{for } ({#1}; {#2}; {#3}) \textbf{do }}

\newcommand{\DONE}{ \textbf{done}}

\newcommand{\FUNCTION}[2]   {\textbf{function}  ${#1}$: \textbf{{#2}}}
\newcommand{\RETURN}[1]     {\textbf{return} {#1}}
\newcommand{\PROC}[1]       {\textbf{procedure} ${#1}$}

\newcommand{\BEGLIST}{\begin{list}{}{\partopsep -3pt \parsep -2pt \listparindent -0pt \labelwidth .5in}}
\newcommand{\ENDLIST}{\end{list}}

\begin{document}

\thispagestyle{empty}
\begin{center}
LaRIA~: Laboratoire de Recherche en Informatique d'Amiens\\
Université de Picardie Jules Verne -- CNRS FRE 2733\\
33, rue Saint Leu, 80039 Amiens cedex 01, France\\
Tel : (+33)[0]3 22 82 88 77\\
Fax : (+33)[0]03 22 82 54 12\\
\underline{http://www.laria.u-picardie.fr}
\end{center}

\vspace{7cm}

\begin{center}
\parbox[t][5.9cm][t]{10cm}
{\center

{\bf Circle Formation of Weak Robots\\ 
     and\\
     Lyndon Words

\bigskip

Yoann Dieudonn\'e$^{\rm a}$ \qquad Franck Petit$^{\rm a}$
}
\bigskip

\textbf{L}aRIA \textbf{R}ESEARCH \textbf{R}EPORT~: LaRIA-2006-05\\
(May 2006)
}
\end{center}

\vfill

\hrule depth 1pt \relax

\medskip

\noindent
$^a$ LaRIA, Université de Picardie Jules Verne, \{Yoann.Dieudonne,Franck.Petit\}@u-picardie.fr

\vspace{-2cm}
\pagebreak

\title{ Circle Formation of Weak Robots and Lyndon Words} 
\author{Yoann Dieudonn\'e \qquad Franck Petit\\
LaRIA, CNRS FRE 2733\\
Universit\'{e} de Picardie Jules Verne\\
Amiens, France}
\date{}
\maketitle

\begin{abstract}
A Lyndon word is a non-empty word strictly smaller in the lexicographic order
than any of its suffixes, except itself and the empty word.  In this paper, we show how Lyndon
words can be used in the distributed control of a set of $n$ weak mobile robots.
By weak, we mean that the robots are anonymous, memoryless, without any common sense of direction,
and unable to communicate in an other way than observation. 
An efficient and simple deterministic protocol to form a regular $n$-gon is presented and proven 
for $n$ prime.  
\end{abstract}

\section{Introduction}
A Lyndon word is a non-empty word strictly smaller in the lexicographic order
than any of its suffixes, except itself and the empty word.  
Lyndon words have been widely studied in the combinatorics of words area~\cite{L83}.  
However, only a few papers consider Lyndon words addressing issues in other 
areas than word algebra, e.g.,~\cite{C04,DR04,SM90}.

In this paper, we address the class of distributed systems where computing units are
\emph{autonomous} \emph{mobile} robots (or agents), i.e., devices equipped 
with sensors which do not depend on a central scheduler and designed to move in a two-dimensional plane. 
Also, we assume that the robots cannot remember any previous observation nor computation performed 
in any previous step.  Such robots are said to be \emph{oblivious} (or \emph{memoryless}). 
The robots are also \emph{uniform} and \emph{anonymous}, i.e, they 
all have the same program using no local parameter (such that an identity) 
allowing to differentiate any of them.
Moreover, none of them share any kind of common coordinate mechanism or common 
sense of direction, and they communicate only by observing the position of the others.


The motivation behind such a weak and unrealistic model is the study of the minimal level of ability 
the robots are required to have in the accomplishment of some basic cooperative tasks in a deterministic 
way~\cite{SS90,SY99,FPSW99,P02}.  
Among them, the \emph{Circle Formation Problem} (CFP) has received a particular attention. 
The CFP consists in the design of a protocol insuring that starting from an initial arbitrary configuration, all $n$ robots 
eventually form a circle with equal spacing between any two adjacent robots.  In other words, a \emph{regular n-gon} must be formed 
when the protocol terminated.  

An informal CFP algorithm is presented in~\cite{D95} to show 
the relationship between the class of pattern formation algorithms and the concept of self-stabilization
in distributed systems~\cite{D00}.  
In~\cite{SS96}, an algorithm based on heuristics is proposed for the formation of a cycle approximation.
A CFP protocol is given in~\cite{SY99} for non-oblivious robots with an unbounded memory.  
Two deterministic algorithms are provided in~\cite{DK02,CMN04}.  In the former work, the robots
asymptotically converge toward a configuration in which they are uniformly distributed on the boundary 
of a circle.  This solution is based on an elegant Voronoi Diagram construction.  
The latter work avoid this construction by making an extra assumption on the initial position of robots.  
All the above solutions work in a semi-asynchronous model.  The solution in~\cite{K05} works on a fully asynchronous model, but 
when $n$ is even, the robots may only achieve a biangular circle---the distance between two adjacent robots is alternatively
either $\alpha$ or $\beta$.  

In this paper, we show a straight application of the properties of Lyndon words.  They are used 
to build and to prove an efficient and simple deterministic protocol solving the CFP 
for a prime number of robots.

\section{Preliminaries}

In this section, we define the distributed system and the problem considered in this paper. 
Next, the Lyndon words are introduced. 

\paragraph{Distributed Model.}

We adopt the model of \cite{SY96}.  
The \emph{distributed system} considered in this paper consists of $n$ robots  
$r_{1}, r_{2},\cdots , r_{n}$, where $n$ is prime---the subscripts $1,\ldots ,n$ are used for notational purpose only.
Each robot $r_{i}$, viewed as a point in the Euclidean plane, move on this two-dimensional 
space unbounded and devoid of any landmark.  When no ambiguity arises, $r_{i}$ also denotes the point in the plane occupied by that robot. 
It is assumed that the robots never collide and that two or more robots may simultaneously occupy the same physical location. 
Any robot can observe, compute and move with infinite decimal precision.
The robots are equipped with sensors allowing to detect the instantaneous position of the other robots in the plane.  
Each robot has its own local coordinate system and unit measure.  
The robots do not agree on the orientation of the axes of their local coordinate system, nor on the unit measure. 
They are \emph{uniform} and \emph{anonymous}, i.e, they all have the same program using no local parameter (such that an identity) 
allowing to differentiate any of them.  
They communicate only by observing the position of the others and they are \emph{oblivious}, i.e.,
none of them can remember any previous observation nor computation performed 
in any previous step. \\

Time is represented as an infinite sequence of time instant $t_0, t_1, \ldots, t_j, \ldots$ 
The set of positions in the plane occupied by the $n$ robots at a given time instant  
$t_j$ ($j\geq0$) is called a \emph{configuration} of the distributed system.   
At each time instant $t_j$ ($j\geq 0$), each robot $r_i$ is either {\it active} or {\it inactive}. 
The former means that, during the computation step $(t_j,t_{j+1})$, using 
a given algorithm, $r_i$ computes in its local coordinate system a position $p_i(t_{j+1})$ depending 
only on the system configuration at $t_j$, and moves towards $p_i(t_{j+1})$---$p_i(t_{j+1})$ can be equal to
$p_i(t_j)$, making the location of $r_i$ unchanged.
In the latter case, $r_i$ does not perform any local computation and remains at the same position. 

The concurrent activation of robots is modeled by 
the interleaving model in which the robot activations are driven by a \emph{fair scheduler}.  
At each instant $t_j$ ($j\geq 0$), the scheduler 
arbitrarily activates a (non empty) set of robots.  
Fairness means that every robot is infinitely often activated by the scheduler.  

\paragraph{The Circle Formation Problem.}
\label{circle}
Consider a configuration at time $t_k$ ($k\geq 0$) in which the positions of the $n$ 
robots are located at distinct positions on the circumference of a 
non degenerate circle $C$---the radius of $C$ is greater than $zero$.
At time $t_k$, the \emph{successor} $r_j$ of any robot $r_i$, $i,j \in 1 \ldots n$ and $i \neq j$, is 
the unique robot such that no robot exists between $r_i$ and $r_j$ on $C$ in the
clockwise direction.  Given $r_i$ and its successor $r_j$ on $C$ centered in $O$,  
$\widehat{r_i O r_j}$ denotes the angle between $r_i$ and $r_j$, i.e., the angle centered in $O$ and 
with sides the half-lines $[O,r_i)$ and $[O,r_j)$.  \\

The problem considered in this paper consists in the design of a 
distributed protocol which arranges a group of $n$ ($n \geq 2$) 
mobile robots with initial distinct positions  
into a \emph{non degenerate regular $n$-gon} 
in finite time
, i.e., the robots eventually 
take place in a non degenerate circle $C$ centered in $O$
such that for every pair $r_i,r_j$ of robots, if $r_j$ is the successor of $r_i$ on $C$,
then $\widehat{r_i O r_j} = \alpha$, where $\alpha= \frac{2\pi}{n}$.  
Angle~$\alpha$ is called the \emph{characteristic angle} of the $n$-gon.

\paragraph{Lyndon Word.}

Let an ordered alphabet $A$ be a finite set of letters.  Denote $\prec$ an order on $A$. 
A non empty \emph{word} $w$ over $A$ is 
a finite sequence of letters $a_1,\ldots,a_i,\ldots,a_l$, $l>0$.  
The \emph{concatenation} of two words $u$ and $v$, denoted $u\circ v$ or simply $uv$, is equal to the word
$a_1,\ldots,a_i,\ldots, a_k,b_1, \ldots, b_j,\ldots,b_l$ such that $u=a_1,\ldots,a_i,\ldots, a_k$ and 
$v=b_1, \ldots, b_j,\ldots,b_l$.  
Let $\epsilon$ be the \emph{empty word} such that for every word $w$, $w\epsilon = \epsilon w = w$.  
The \emph{length} of a word $w$, denoted by $|w|$, is equal to the number of letters of $w$---$|\epsilon|=0$.  

A word $u$ is \emph{lexicographically} smaller than or equal to a word $v$, 
denoted $u \preceq v$, iff there exists either a word $w$ such that $v=uw$ or 
three words $r,s,t$ and two letters $a,b$ 
such that $u=ras$, $v=rbt$, and $a\prec b$.
%

Let $k$ and $j$ be two positive integers.  The {\it $k^{th}$ power} of a word $w$ is 
the word denoted $s^k$ such that $s^0 = \epsilon$, and $s^k = s^{k-1} s$.
A word $u$ is said to be \emph{primitive} if and only if $u = v^k \Rightarrow k=1$. 
The \emph{$j^{th}$ rotation} of a word $w$, notation $R_j (w)$, is defined by:

$$ 
R_j (w) \stackrel{\mathrm{def}}{=}
\left\{
  \begin{array}{ll}
     \epsilon & \mbox{if } w = \epsilon \\
     a_j,\ldots, a_l,a_1,\ldots,a_{j-1} & \mbox{otherwise } ( w = a_1,\ldots,a_l,\ l\geq 1 )
  \end{array}
\right.
$$
    
Note that $R_1 (w) = w$. A word $w$ is said to be {\it minimal} iff $\forall j \in 1,\ldots, l$,
$w \preceq R_j (w)$.
%

\begin{definition}[Lyndon Word]
\label{def:LW}
A word $w$ ($|w|>0$) is a Lyndon word iff $w$ is nonempty, primitive and minimal, i.e.,
$w \neq \epsilon$ and $\forall j \in 2,\ldots, |w|,\ w \prec  R_{j}(w)$.
\end{definition}
For instance, if $A=\{a,b\}$, then $a$, $b$, $ab$, $aab$, $abb$ are Lyndon words, whereas $aba$, and $abab$ are not---
$aba$ is not minimal ($aab \preceq aba$) and $abab$ is not primitive ($abab = (ab)^2$).

\section{Algorithm}

In this section, we present our main result based on Lyndon words 
for $n$ robots, $n$ being prime and greater than or equal to $5$.  
Note that if the system contains two robots only, then CFP is trivially 
solved because they always form a $2$-gon.  
If $n=3$, then the three robots always form a triangle, which must be equilateral to
form a regular $3$-gon.  If the triangle is not equilateral, then either
($1$) the three robots belongs to the same line, ($2$) they form an isosceles triangle, or 
($3$) they form an ordinary triangle.  In all cases, it is always possible to elect a unique
robot as the leader which moves to the unique position making the triangle equilateral---
%
a formal algorithm for $n=3$ is given and proved in the appendix.  

The technique developed in this paper is based on a Leader Election using properties of Lyndon words.  
The words we consider are made over the alphabet such that the letters are the angles between 
neighboring robots located on the boundary of a unique non degenerate circle. 
So, we focus only on configurations where the robots are either all or all but one on the boundary
of the circle.  
Thus, we integrate the first algorithm proposed in \cite{DK02} in our solution, in the sequel refered to 
Algorithm~$\phi_{circle}$.  

\begin{theorem}[\cite{DK02}]
\label{th:DK}
Starting from an arbitrary configuration where $n$ robots are located at distinct positions, 
Algorithm~$\phi_{circle}$ leads the robots into a configuration where the robots are located on 
the boundary of a non degenerate circle.  
\end{theorem}

In the rest of this section, 
we first present how Lyndon words are used to make a leader election
among $n$ robots located on the boundary of a unique circle.  
Next, we define a particular type of configurations called \emph{oriented} configurations.  
We then give an algorithm to arrange the robots in a regular $n$-gon starting
from an oriented configuration.  
Finally, we provide our general scheme to solve the Circle Formation Problem.

\subsection{Leader Election}
\label{sub:LE}

In this subsection,
we use the subscript $i$ in the notation of a robot $r_i$, $i \in 1\ldots n$, to denote the order of the robots in an
arbitrary clockwise direction on $C$.  We proceed as follows:
A robot is arbitrarily chosen as $r_1$ on $C$.
Next, for any $i \in 1\ldots n-1$, $r_{i+1}$ denotes the successor of $r_i$ on $C$ (in the
clockwise direction).  Finally, the successor of $r_n$ is $r_1$. 

\begin{figure}[!htbp]
\begin{center}
    \epsfig{file=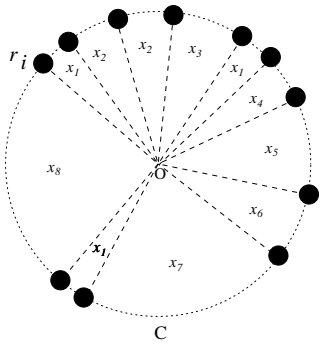, width=0.25\linewidth}
\end{center}
\caption{$SA(r_i)= x_1  (x_2)^2 x_3  x_1  x_4  x_5  x_6  x_7  x_1  x_8$.}
\label{fig:SA}
\end{figure}

Let the alphabet $A$ be the set of $k$ ($k\leq n$) strictly positive reals 
$x_1,x_2,\ldots, x_k$ such that 
$\forall i \in 1\ldots n$, there exists $j \in 1\ldots k$
such that $x_j=\widehat{r_i O r_{i+1}}$, where $O$ is the center of $C$.
An example of such an alphabet is shown in Figure~\ref{fig:SA}---$A= \{x_1,x_2,x_3,x_4,x_5,x_6,x_7,x_8\}$.

The order on $A$ is the natural order ($<$) on the reals.  So, the lexicographic order $\preceq$
on the words made over $A$ is defined as follows:
$$
u\preceq v \stackrel{\mathrm{def}}{\equiv} 
\left( \exists w |\ v=uw \right) \vee
\left( \exists r,s,t,\ \exists a,b \in A |\ (u=ras) \wedge (v=rbt) \wedge (a<b) \right)
$$
For instance, if $A = \{\frac{\pi}{3}, \frac{\pi}{2}\}$, then 
$
  (\frac{\pi}{3}) \preceq (\frac{\pi}{3} \frac{\pi}{3}) \preceq 
  (\frac{\pi}{3} \frac{\pi}{2}) \preceq  (\frac{\pi}{3} \frac{\pi}{2} \frac{\pi}{2})
  \preceq (\frac{\pi}{2})$.

For each robots $r_i$, let us define the word $SA(r_i)$ (respectively, $\overline{SA(r_i)}$) 
over $A$ ($SA$ stands for ``string of angles''~\cite{CP02}) as follows:
$$
\begin{array}{c}
  SA(r_i)=x_i x_{i+1} \ldots x_n x_1 \ldots x_{i-1}\\
  \mbox{(resp. } \overline{SA(r_i)}=x_{i-1} \ldots x_1 x_n x_{n-1} \ldots x_i \mbox{)}
\end{array}
$$

An example showing a string of angle is given in Figure~\ref{fig:SA}.
Note that for every robot $r_i$, $|SA(r_i)| = |\overline{SA(r_i)}| = n$.  Moreover, if the configuration is a regular $n$-gon, then 
for every robot $r_i$, $SA(r_i)= \overline{SA(r_i)} = \alpha^{n}$, 
where $\alpha=\frac{2\pi}{n}$ is the characteristic angle of the $n$-gon.

\begin{lemma}
\label{lem:LWS}
If all the robots are at distinct positions on the boundary of a same circle $C$ without forming 
a regular $n$-gon, then there exists exactly one robot $r_i$ such that $SA(r_i)$ is a Lyndon Word.
\end{lemma}
\begin{proof}
We prove this in two steps.  First, we show that $r_i$ always exists.  Next, we show the uniqueness of $r_i$.
\\

\noindent \emph{Existence}.  Let $minSA$ be the string of angle such that 
$\forall i \in 1\ldots n,\ minSA \preceq SA(r_i)$.  Let us show that $minSA$ is a Lyndon word.  
By definition, $minSA$ is minimal.
Assume by contradiction that $minSA$ is not primitive. 
So, by definition, there exists a word $u$ such as $minSA=u^k$ with $k>1$.  
By definition of the $k^{\mbox{th}}$ power of a word $u$, $|minSA|=k*|u|$, $|u|$ 
being a divisor of $n$.  Since $|minSA|=n$ and $n$ is prime, there are only two cases to consider:
\begin{enumerate}
\item $|u| = |minSA|$.  Since $|minSA|=k*|u|$, this implies that $k=1$, which contradicts $k>1$.
\item $|u| = 1$.  In this case, $u$ is a letter ($\in A$), and $minSA=u^n$.  Since the letters 
are angles, the angle between every pair of successive robots is 
equal to the value $u$.  So, the robots form a regular $n$-gon and $u = \alpha$, the characteristic angle 
of the $n$-gon.  This contradicts that the configuration is not a regular $n$-gon.
\end{enumerate}

\noindent \emph{Unicity}.  Assume by contradiction that there exists two different robots 
$r_i,r_j$ ($r_i\neq r_j$) such that $SA(r_i)$ and $SA(r_j)$ are Lyndon words.  
Since $SA(r_j)$ (respectively, $SA(r_j)$) is a Lyndon word, by Definition~\ref{def:LW},
$\forall k \in 2\ldots n,\ SA(r_i) \prec R_k(SA(r_i))$ 
(resp. $\forall k' \in 2\ldots n,\ SA(r_j) \prec R_k(SA(r_j))$).  
Since $r_i \neq r_j$, there exists $k \in 2\ldots n$ such that $SA(r_j) = R_k(SA(r_i))$ (resp. 
there exists $k' \in 2\ldots n$ such that $SA(r_i) = R_{k'}(SA(r_j))$).
Hence, $SA(r_i) \prec SA(r_j)$ and $SA(r_j) \prec SA(r_i)$.  A contradiction.
\end{proof}

Clearly, following the reasoning as for Lemma~\ref{lem:LWS}:
\begin{lemma}
\label{lem:LWSb}
If all the robots are at distinct positions on the boundary of a same circle $C$ without forming a 
regular $n$-gon, then there exists exactly one robot $r_i$ such that $\overline{SA(r_i)}$ is a Lyndon Word.
\end{lemma}

Let $LWS$ be the set of robots $r_i$ such that $SA(r_i)$ or $\overline{SA(r_i)}$ is a Lyndon word.  

\begin{lemma}
\label{lem:LWS2}
If all the robots are at distinct positions on the boundary of a same circle $C$ without forming a 
regular $n$-gon, then $|LWS| = 2$.
\end{lemma}
\begin{proof}
 From Lemmas~\ref{lem:LWS} and~\ref{lem:LWSb}, 
there exists exactly one robot $r_a$ such that $SA(r_a)$ is a Lyndon word, and 
exactly one robot $r_b$ such that $\overline{SA(r_b)}$ is a Lyndon word.  
However, $r_a$ can be the same robot as $r_b$.  So, $1\leq |LWS| \leq 2$.

Let assume that $|LWS|=1$, i.e., there exists a unique robot $r$ such that both 
$SA(r)=a_1 a_2 \ldots a_n$ and $\overline{SA(r)}=a_n a_{n-1} \ldots a_1$ are Lyndon words. 
Since the robots do not form a regular $n$-gon, $SA(r) \neq a_1^n$.  So, $SA(r)$ contains at least
two different letters.  There are three cases to consider:
\begin{enumerate}
\item $a_n \prec a_1$.  So, $a_n a_1 a_2 \ldots a_{n-1} \prec a_1  \ldots  a_{n-1} a_n$, i.e., 
$R_n(SA(r)) \prec SA(r)$.  This contradicts that $SA(r)$ is a Lyndon word---by 
Definition~\ref{def:LW}, $SA(r) \prec R_n(SA(r))$.  
\item $a_1 \prec a_n$.  So, $a_1 a_n a_{n-1} \ldots a_2 \prec a_n  a_{n-1}  \ldots  a_2 a_1$, i.e., 
$R_n(\overline{SA(r)}) \prec \overline{SA(r)}$.  This contradicts that $\overline{SA(r)}$ is 
a Lyndon word---by Definition~\ref{def:LW}, $\overline{SA(r)} \prec \overline{R_n(SA(r))}$.  
\item $a_1 = a_n$.  Since $SA(r)$ is a Lyndon word, by Definition~\ref{def:LW}, we have 
$a_1 a_2  \ldots a_{n-1} a_n \prec a_n  a_1  a_2  \ldots  a_{n-1}$. 
Since $a_1 = a_n$, we have $a_2  \ldots a_{n-1} a_n \prec a_1  a_2  \ldots  a_{n-1}$.
Since $a_1 = a_n$ again, we have also 
$a_2  \ldots a_{n-1} a_n a_1 \prec a_1  a_2  \ldots  a_{n-1} a_n$, i.e., 
$R_2(SA(r)) \prec SA(r)$.  
This contradicts Definition~\ref{def:LW}, $SA(r) \prec R_2(SA(r))$.  
\end{enumerate}
\end{proof}

Now, consider the distributed computation of string of angles.  
We borrow Algorithm~\ref{algo:SA} from \cite{CP02} which describes a function called Function~$ComputeSA$.  
Each robot $r_i$ arbitrarily determines its own clockwise direction 
of $C$ in its local coordinate system.  Note that since the robots are 
uniform, all of them apply the same algorithm to determine the clockwise direction.  However, the robots do not
share a common coordinate system.  So, for any pair $r_i,r_j$, the clockwise direction of $r_i$ (resp., $r_j$) 
may be the counterclockwise of $r_j$ (resp., $r_i$).  

\begin{algo}[htb]
\begin{small}
\begin{tabbing}
  xxxxx \= xxxxx \= xxxxx \= xxxxx \= xxxxx \= xxxxx \= xxxxx \= xxxxx \= xxxxx \= \kill 
\FUNCTION{ComputeSA(r_j)}{word}\\
\>  $SA:=\epsilon$;\ $r:=r_j$;\\
\>  \FOR {$k=1$}{$k\leq n$}{$k:=k+1$}\\
\>  \> $r':= Succ(r)$; 
$SA := SA \circ Angle(r,r')$; 
$r:=r'$;\\
\>  \DONE\\
\>  \RETURN{$SA$};
\end{tabbing}
\end{small}
\caption{Function $ComputeSA$ for any robot $r_i$}
\label{algo:SA}
\end{algo}


Algorithm~\ref{algo:SA} uses two functions, $Succ(r)$ and $Angle(r,r')$.  The former returns the successor of $r$
in the local coordinate system of the robot executing $Succ(r)$.
The latter returns the absolute value of $\widehat{rOr'}$, where $O$ is the center of $C$.  Using this algorithm,
each robot $r_i$ computes $SA(r_j)$, $\forall j \in 1\ldots n$.  
Similarly, for each robot $r_j$, $\overline{SA(r_j)}$ can also be easily computed by any $r_i$ by either 
computing the mirroring word of $SA(r_j)$ or following 
Algorithm~\ref{algo:SA} and replacing Function~$Succ(r)$ by a function $Pred(r)$, which 
returns the unique predecessor of $r$ in the counterclockwise direction.\\ 

\begin{figure}[!htbp]
\begin{center}
    \epsfig{file=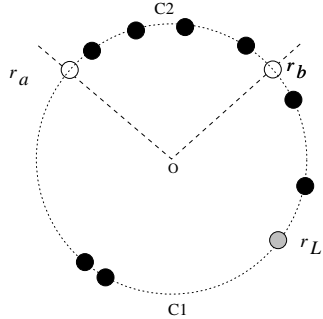, width=0.25\linewidth}
\end{center}
\caption{An example showing how the leader robot $r_L$ is computed ($n=11$).}
\label{fig:LE}
\end{figure}

Let us now describe how the above results can be used to elect a leader.  
An example showing our method is given in Figure~\ref{fig:LE} with $n=11$. 
Following Lemmas~\ref{lem:LWS}, \ref{lem:LWSb}, and \ref{lem:LWS2}, 
when any robot computes the set $LWS$, then $LWS = \{r_a,r_b\}$ and $r_a$ and $r_b$ refer to
the same robots for every robot. 
Consider both half-lines $[O,r_a)$ and $[O,r_b)$.
These two half-lines divide the circle $C$ into two sides,
$C1$ and $C2$, where $n_{C1}$ (resp. $n_{C2}$) represent the number of robots inside $C1$ (resp $C2$).
Note that $n_{C1} + n_{C2} = n-2$.  Since $n-2$ is odd ($n$ is prime), there exists one side with an
even number of robots, and one side with an odd number of robots.  Without lost of generality,
let us assume that $n_{C1}$ is odd.  Let $r_L$ be the unique robot which is the median robot on $C_{1}$,
i.e., the $(\lfloor \frac{n_{C1}}{2} \rfloor +1)^{\mbox{th}}$ robot starting indifferently from $r_a$ or $r_b$.

Let us define Function $Elect()$ which returns the unique leader robot $r_L$ for every robot $r_i$.
\begin{lemma}
\label{lem:LE}
If all the robots are at distinct positions on the boundary of a same circle $C$ without forming a 
regular $n$-gon, then for every robot $r_i$, Function~$Elect()$ returns a unique leader robot $r_L$ 
among the $n$ robots.  
\end{lemma}

\subsection{Oriented Configuration}
\label{sub:OC}

A configuration is said to be {\it oriented} 
if the following conditions hold:
\begin{enumerate}
\item All the robots are at distinct positions on the same circle $C_\OR$, except only one of them, 
called $r_\OR$, located inside $C_\OR$;
\item $r_\OR$ is not located at the center $O$ of $C_\OR$;
\item there is no robot on $C_\OR \cap [O,r_\OR)$. 
\end{enumerate}

\begin{figure}[!htbp]
\begin{center}
    \epsfig{file=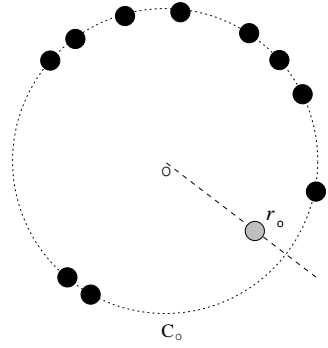, width=0.25\linewidth}
\end{center}
\caption{An oriented configuration ($n=11$).}
\label{fig:OC}
\end{figure}

An example of an oriented configuration is shown in Figure~\ref{fig:OC} with $n=11$. 
Let us denote an oriented configuration by the pair of its two main parameters, i.e.,
$(C_\OR,r_\OR)$.  Two oriented configurations $(C^\alpha_\OR,r^\alpha_\OR)$ and 
$(C^\beta_\OR,r^\beta_\OR)$ are said to be \emph{equivalent} if
$C^\alpha_\OR = C^\beta_\OR$ and both $r^\alpha_\OR$ and $r^\beta_\OR$ are located 
at the same position.  In other words, the only possible 
difference between two equivalent 
oriented configurations $(C^\alpha_\OR,r^\alpha_\OR)$ and $(C^\beta_\OR,r^\beta_\OR)$ 
is different positions of robots between $C^\alpha_\OR$ and $C^\beta_\OR$.

We now describe Procedure~$\phi_{\OR}$ shown in Algorithm~\ref{algo:OC}.  This procedure assumes that
the configuration is oriented.  In such a configuration, we will build a partial order among the robots
to eventually form an $n$-gon.

\begin{algo}[htb]
\begin{small}
\begin{tabbing}
  xxxxx \= xxxxx \= xxxxx \= xxxxx \= xxxxx \= xxxxx \= xxxxx \= xxxxx \= xxxxx \= \kill 
\PROC{\phi_{\OR}}\\
\> $C_\OR := \mbox{circle where } n-1 \mbox{ robots are located}$;\\
\> $O:= \mbox{center of }C_\OR$;\\
\> $r_\OR := \mbox{robot inside } C_\OR$;\\
\> $p_1 := C_\OR \cap [O,r_\OR)$;\\ 
\> $PS := FindFinalPos(C_\OR ,p_1)$;\\
\> $FRS := \mbox{set of robots which are not located on a position in } PS \mbox{, except } r_\OR$;\\
\> \IF{$FRS = \emptyset$}\\
\> \THEN \> // Every robot on $C_\OR$ is located on a final position ($\in PS$).\\
\>       \> \IF{$r_i = r_{\OR}$} 
             \THEN move to Position $p_1$;\\
\>       \> \ENDIF\\
\> \ELSE \> $EFR := ElectFreeRobots(FRS)$;\\
\>       \> \IF{$r_i \in EFR$}
            \THEN move to Position $Associate(r_i)$; 
\end{tabbing}
\end{small}
\caption{Procedure $\phi_\OR$ for any robot $r_i$ in an oriented configuration if $n\geq 5$}
\label{algo:OC}
\end{algo}

Let $p_1,\ldots,p_n$ be the final positions of the robots when the regular $n$-gon is formed.  
Let $p_1 = C_\OR \cap [O,r_\OR)$. Then,  for each $k \in 2\ldots n$, $p_k$ is the 
point on $C_\OR$ such that $\widehat{p_1 O p_k} = \frac{2k\pi}{n}$. 
Clearly, while the distributed system remains in an equivalent oriented configuration, all the final 
position remain unchanged for every robot. 
A position $p_k$,  $k \in 2\ldots n$, is said to be \emph{free} if no robot takes place at $p_k$.  Similarly, 
a robot $r_i$ on $C_\OR$ (i.e., $r_i  \neq r_\OR$) is called a \emph{free} robot if its current position 
does not belong to $\{p_2, \ldots, p_n\}$.
Define Function $FindFinalPos(C_\OR,p_1)$ which returns the set of final positions on $C_\OR$ 
with respect to $p_{1}$.  Clearly, all the robots compute the same set of final positions, stored in $PS$.  
Each robot also temporarily stores the set of free robots in the variable called $FRS$.  

Basically, if $FRS = \emptyset$ while the $n$-gon is not formed, then it remains $r_\OR$ only 
to move from its current position inside $C_\OR$ to $p_1$.  Otherwise, the robots move in waves
to the final positions following the order defined by Function $ElectFreeRobots()$.  The elected 
robots are the closest free robots from $p_1$.  
Clearly, the result of Function $ElectFreeRobots()$ return the same set of robots for 
every robot.  Also, the number of elected robots is at most
equal to $2$, one for each direction on $C_\OR$ with respect to $p_1$.  Note that it can be 
equal to $1$ when there is only one free robot, i.e., when only one robot on $C_\OR$ did
not reach the last free position.

Function $Associate(r)$ assigns a unique free position to an elected robot as follows:\newline
\noindent If $ElectFreeRobots()$ returns only one robot $r$, then $r$ is associated to the unique free
remaining position $p$.  This allows $r$ to move to $p$.  
If $ElectFreeRobots()$ returns a pair of robots $\{r_a,r_b\}$ ($r_a \ne r_b$), then the closest robot 
to $p_1$, in the clockwise (respectively, counterclockwise) is associated with the closest position to 
$p_1$ in the clockwise (resp., counterclockwise) direction.  
Note that, even if the robots may have opposite clockwise directions, $r_a$, $r_b$, and their associated 
positions are the same for every robot. 

\begin{lemma}
\label{lem:oc}
If the robots are in an oriented configuration at time $t_j$ ($j\geq0$), then
at time $t_{j+1}$, either the robots are in an equivalent oriented configuration 
or they form a regular $n$-gon.
\end{lemma}
\begin{proof}
By contradiction, assume that starting from an oriented configuration at time $t_j$,
the robots are not in an equivalent oriented configuration and they 
do not form a regular $n$-gon at time $t_{j+1}$.  By assumption, at each time instant $t_j$,
at least one robot is active.  So, by fairness, starting from an oriented 
configuration, at least one robot executes Procedure~$\phi_{\OR}$.  Assume first that no 
robot executing Procedure~$\phi_{\OR}$ moves from $t_j$ to $t_{j+1}$.  In that case,
since the robots are located on the same positions at $t_j$ and at $t_{j+1}$, the robots 
are in the same oriented configuration at $t_{j+1}$. An oriented configuration 
being equivalent to itself, this contradicts the assumption. 
So, at least one robot moves from $t_j$ to $t_{j+1}$.  
There are two cases to consider:
\begin{enumerate}
\item  No final position is free at time $t$.  So, every robots on $C_{\OR}$ ($\neq r_\OR$) 
is located on a final position and no final position is free. 
Then, for every robot executing Procedure~$\phi_{\OR}$, $FRS= \emptyset$.  
In that case, only $r_\OR$ is allowed to move.  Since by assumption
at least one robot moves from $t_j$ to $t_{j+1}$, 
$r_\OR$ moves from its current position to $p_1$.  Therefore, the robots form a regular $n$-gon
at time $t_{j+1}$.  A contradiction.
\item Final positions are free at time $t$.  Then, for every robot executing Procedure~$\phi_{\OR}$, 
$FRS\neq \emptyset$.  By construction, there exists either one or two robots which belongs 
to the set $EFR$ and $r_\OR$ is not allowed to move.  
Therefore, at least one of them moves from its current position on $C_\OR$ to its associated 
free position.
Obviously, the robots are again in distinct positions from $t_j$ to $t_{j+1}$. Furthermore,
 since all the final positions are on $C_\OR$ and $r_\OR$ remains at the same position
from $t_j$ to $t_{j+1}$, the robots are in an equivalent oriented configuration at $t_{j+1}$.  
A contradiction.
\end{enumerate}
\end{proof}


\begin{theorem}
\label{th:oc}
Starting from an oriented configuration, Algorithm $\phi_{oriented}$ solves the Circle Formation Problem.
\end{theorem}
\begin{proof}
It follows from the proof of Lemma~\ref{lem:oc} that in a oriented configuration, there exists 
at least one robot which can move to any final position.  The theorem follows by fairness 
and Lemma~\ref{lem:oc}.
\end{proof}

\subsection{Main Algorithm} 

The main part of our solution is presented in Algorithm~\ref{algo:ngon}.  
Using Algorithm $\phi_{circle}$~\cite{DK02}, the robots are first placed in a circle $C$.  Then, 
a robot is pointed out as the leader $r_\OR$---using Function~$Elect()$ described in Subsection~\ref{sub:LE}.  
Next, Robot~$r_\OR$ moves inside the circle on the segment $[O,r_\OR]$, at a position arbitrarily chosen 
at the middle of $[O,r_\OR]$.  Finally, following the partial order provided by Procedure~$\phi_{\OR}$,
the robots eventually take place in a regular $n$-gon. 

\begin{algo}[htb]
\begin{small}
\begin{tabbing}
  xxxxx \= xxxxx \= xxxxx \= xxxxx \= xxxxx \= xxxxx \= xxxxx \= xxxxx \= xxxxx \= \kill 
\IF{the robots do not take place in a regular $n$-gon} \\
\THEN \> \IF{the robots are in an oriented configuration}\\
      \> \THEN \> Execute Procedure $\phi_{\OR}$\\
      \> \ELSE \> \IF{the robots take place in a circle $C$}\\
      \>       \> \THEN \> \IF{$r_i = Elect()$} //Robot $r_i$ is the leader robot\\
      \>       \>       \> \THEN \> move to Position $p$ such that $p = \frac{Dist(r_i,O)}{2}$, $O$ being the center of $C$;\\
      \>       \>       \> \ENDIF\\
      \>       \> \ELSE \> Execute Procedure $\phi_{circle}$;
\end{tabbing}
\end{small}
\caption{Procedure $\phi_\OR$ for any robot $r_i$ in an oriented configuration if $n\geq 5$}
\caption{($\phi_{n\mbox{-gon}}$) Algorithm for any robot $r_i$ if $n\geq 5$}
\label{algo:ngon}
\end{algo}

\begin{theorem}
\label{th:ngon}
Algorithm $\phi_{n\mbox{-gon}}$ solves the problem of circle formation for $n\geq 5$ robots.
\end{theorem}
\begin{proof}
 From Algorithm~\ref{algo:ngon}, if the robots form an $n$-gon, then no robot moves and the 
the robots form an $n$-gon forever.  If the robots do not form an $n$-gon, then 
from Theorem~\ref{th:DK}, Algorithm~\ref{algo:ngon} again, Lemma~\ref{lem:LE}, and 
Theorem~\ref{th:oc}, the robots eventually form an $n$-gon.
\end{proof}

\section{Conclusion}
We showed how Lyndon words can be used in the distributed control of a set of $n$ anonymous robots being
memoryless, without any common sense of direction, and unable to communicate in an other way than observation. 
An efficient and simple deterministic protocol to form a regular $n$-gon was presented for
a prime number of robots.  We believe that the new idea presented in this paper should help in the 
design of the protocol for the Circle Formation Problem with any arbitrary number of robots.  
That would be the main goal of our future research.

\section*{Acknowledgements}
We are grateful to Gw\'ena\"el Richomme for the valuable discussions.

\begin{small}
\bibliographystyle{alpha}
\bibliography{prime}
\end{small}

\newpage
\appendix
\section{Appendix : Circle Formation for $3$ Robots}
The aim of the algorithm, shown in Algorithm~\ref{algo:3gon}, is to lead the $3$ robots to eventually
take place in a regular $3$-gon, i.e., an equilateral triangle.  
The correctness proof of Algorithm~\ref{algo:3gon} is based on the following lemma: 
\begin{lemma}
\label{lem:3gon}
If the three robots $\{r_1,r_2,r_3\}$ do not form an equilateral triangle, then it is possible to elect only one robot among them.
\end{lemma}
\begin{proof}
If $\{r_1,r_2,r_3\}$ do not form 
an equilateral triangle, then either (Case~$1$) they belongs to the same line, (Case~$2$) they form an isosceles 
triangle, or (Case~$3$) they form an ordinary triangle.  In the first case, the elected robot is 
the median one, which is unique.  In Case~$2$, the elected robot is the one placed at the unique angle different
from the two others.  In Case~$3$, there exists a unique angle which is strictly smallest than the others.  The 
elected robots is the one placed at this unique angle. 
\end{proof}

Lemma~\ref{lem:3gon} allows to define the Boolean function $Elect()$, which can be executed by any robots $r_i$.  
$Elect()$ returns the unique leader robot according to Lemma~\ref{lem:3gon}.

\begin{algo}[htb]
\begin{small}
\begin{tabbing}
  xxxxx \= xxxxx \= xxxxx \= xxxxx \= xxxxx \= xxxxx \= xxxxx \= xxxxx \= xxxxx \= \kill 
\IF{the robots do not take place in a regular $3$-gon} \\
\THEN \> \IF{$r_i = Elect()$}\\
      \> \THEN \> $(r_j,r_k)$ denotes the two other robots than myself;\\
      \>       \> move to Position $p$ such that $\{p,r_j,r_k\}$ form a regular $3$-gon;
\end{tabbing}
\end{small}
\caption{($\phi_{3\mbox{-gon}}$) Algorithm for any robot $r_i$ if $n=3$}
\label{algo:3gon}
\end{algo}

\begin{theorem}
\label{theorem1}
Algorithm $\phi_{3\mbox{-gon}}$ solves the problem of circle formation for three robots.
\end{theorem}
\begin{proof}
 From Lemma \ref{lem:3gon}, we can distinguish a unique robot, called the leader.
According to Algorithm~\ref{algo:3gon}, if the robots do not form an equilateral triangle at time $t$, then only 
the leader is allowed to move.  Since by assumption at least one robot is activated at each instant time, 
at time $t+1$ the robots form an equilateral triangle. 
\end{proof}

\end{document}